\documentstyle[12pt]{article}
\newcommand{\beq}{\begin{equation}}
\newcommand{\eeq}{\end{equation}}
\begin{document}
\topmargin -1.8cm
\oddsidemargin -0.8cm
\evensidemargin -0.8cm

\title{\Large{Partition Function Zeros of an Ising Spin Glass}}
\vspace*{1cm}

\author{
{\sc P.H. Damgaard}\\CERN -- Geneva\\ ~\\and\\~\\ {\sc J. Lacki} \\
Universit\'{e} de Gen\`{e}ve\\32 Quai Ernest Ansermet\\CH-1211 Geneva
4, Switzerland}
\maketitle
\vfill
\begin{abstract}
We study the pattern of zeros emerging from
exact partition function evaluations of Ising spin glasses on
conventional finite
lattices of varying sizes. A large number of random bond
configurations are probed in the framework of quenched averages.
This study is motivated by the relationship between hierarchical
lattice models whose partition function zeros fall on Julia sets
and chaotic renormalization group flows in such models with
frustration, and by the possible connection of the latter with spin
glass behaviour.
In any finite volume, the simultaneous distribution of the zeros of all
partition functions can be viewed as part of the more general problem
of finding the
location of all the zeros of a certain class of random polynomials with
positive integer coefficients. Some aspects of this problem have been studied
in various areas of mathematics, and we show in particular how polynomial
mappings which are used in graph theory to classify graphs, may help in
characterizing the distribution of zeros. We finally discuss the
possible limiting set of these zeros as the volume is sent to infinity.

\end{abstract}
\vfill
\vspace{20mm}
\begin{flushleft}
CERN--TH-7383/94 \\
July 1994 \\
\end{flushleft}
\newpage


\section{Introduction}

The $\pm K$ Ising spin glass is defined as a nearest-neighbour
Ising model with the bonds
$K_{ij}$ (between sites $i$ and $j$) distributed randomly on the lattice
according to a given probability distribution. The nature of the possible
phase transitions of such a model in different number of dimensions is,
despite its apparently simple
structure, still to a large extent an open question (see, $e.g.$, ref.
\cite{review} for some excellent recent reviews). But also the physics
of the ordered or disordered spin glass phases themselves needs to
be better understood. Although models of the Ising spin glass kind are
easy to define formally, finding even approximate analytic solutions is
strikingly difficult. And just as the highly complex dynamics of these
models makes it hard to apply conventional analytical approaches, this
very aspect is also making standard Monte Carlo simulations almost
prohibitively difficult for systems of sufficiently large volume
\cite{MC}. The
combination of both analytical and numerical problems when studying these
models thus seems to call for a new approach, one that combines
the exactness of analytical treatments with the computational power of
numerical evaluations. One very interesting combined approach of this
kind is the {\em exact evaluation of finite-volume spin glass partition
functions}\cite{Bhanot,Saul}, based on either variations of the numerical
transfer-matrix method or, specific to two dimensions, exact rewrites
of the full high-temperature expansion. With exact (numerical, but with as high
precision as needed, without statistical or systematic errors) computation
of Ising spin glass partition functions the apparently only remaining problem
in ``solving'' the theory is to understand how the results scale as the finite
volumes are taken to infinity.\footnote{In practice, however, the limitations
of the procedure show up earlier, in the restriction of the partition function
evaluation to only a subset of the full set of random bond configurations
on the finite-volume lattice. See below.}

The purpose of the present paper is to discuss the
distribution of partition-function zeros of a $\pm K$ Ising spin glass with
nearest neighbour interaction in the complex temperature plane.
This is a problem for which the method of
exact finite-volume partition function evaluation is the only
one applicable,
and we will use it as the basis of our analysis. With present-day
numerical routines, the location of the
partition-function zeros can be found to any practically required
precision for systems
of very large volumes. In this location of the partition-function zeros
is hidden a wealth of information about the model under study. This has been
known since the study of Ising model zeros in both the complex
activity plane \cite{Lee} (the so-called Lee-Yang zeros), and, more relevant
for the present discussion, the complex temperature plane \cite{Fisher}
(now known as Fisher zeros). Of particular interest are of course those
partition-function zeros that lie close to possible phase transition points or
phase transition boundaries for physical, real, values of the magnetic field
or temperature. Here renormalization group (RG) considerations
or finite-size scaling theory give direct
connections to universal critical properties of the system
\cite{Fisher1,Itzykson}. But, as we will argue below, at least for the
case of Ising spin glasses, the location of partition function zeros
{\em away} from critical points may also provide non-trivial information about
the dynamics of the system.

Our original motivation for the present study was precisely to use
the overall location of partition function zeros to shed new light on
the nature of spin glass
interactions at arbitrary points in the phase diagram as, for example,
real-space renormalization group transformations
are taking us to the long-distance limit. This approach was inspired
in an indirect way by
the very interesting suggestion of McKay, Berker and Kirkpatrick \cite{chaos}
that an
exactly solvable ``frustrated'' hierarchical lattice model, with what
turns out to be chaotic
real-space RG transformations, describes a spin glass.
Intuitively this proposal is supported by the fact that under chaotic RG
transformations the typical behavior is an essentially random jump between
strong and weak coupling as larger and larger distance scales are probed,
a behaviour suspiciously reminiscent of a spin glass phase. Other
circumstantial evidence
also suggests a connection to chaotic phenomena \cite{Bray}, but difficulties
in constructing reliable real-space RG transformations for spin glasses on
Bravais lattices\footnote{Chaotic behavior may of course show up as
artifacts of approximate RG transformations, without any physical significance
\cite{Thorleifsson}. In two dimensions, a regular renormalization group flow
between fixed points of unitary, Poincar\'{e} invariant theories is
ensured by Zamolodchikov's $c$-theorem \cite{Zamo}. Spin glasses models can
evade this theorem, since they do not satisfy the assumption of translation
invariance. For a discussion of related issues, and some measurable
predictions of chaotic RG flows, see ref. \cite{me}.}
makes it difficult to establish a direct connection. For example, does spin
glass behavior automatically entail chaotic real-space RG flows? Is the
converse true?

As we shall suggest in this paper, there may be an indirect way of assessing
the possible
connection between chaotic RG flows and spin-glass behaviour. One striking
consequence of the exact (chaotic)
renormalization group transformations on hierarchical lattices is the limiting
distribution of partition function zeros on, in general, fractal Julia sets
\cite{Derrida0}. If ordinary Bravais-lattice spin glass models are
describable by similar chaotic RG transformations, an interesting
possibility is that the fractal nature of the set of partition
function zeros might be preserved. Certainly, {\em if} partition
function zeros of certain spin-glass models turned out to form
fractal sets in the thermodynamic limit, this would be a strong
hint that the dynamics behind these models may be related to chaotic RG
transformations. Can the argument be strengthened? The fact that hierarchical
lattice models give rise to (in general, but not always, fractal) Julia
sets for the partition function zeros is more general, and not tied directly
to spin glass behaviour of the model under study. (Interestingly,
partition function zeros on Julia sets can also arise due to approximations
of exact real-space renormalization group transformations. For an
illustration of this in the context of the 2-d Ising model, see ref.
\cite{Derrida}.) But the argument does run the other way: hierarchical
models with frustration give spin glass behaviour, do have chaotic
renormalization group transformations, and do indeed lead to partition function
zeros lying on (in general fractal) Julia sets. One of the physical
consequences of chaotic renormalization group transformations is that in
general the free energy will have an infinite sequence of singularities
\cite{Derrida1}. This behaviour is again compatible with a scenario of
partition function zeros falling on a fractal set in the complex temperature
plane. We will discuss these issues in more detail in section 2.

Although the possibility of observing fractal behavior in the distribution
of the partition function zeros served as the original motivation for the
present work, we have in the process of investigation uncovered a number
of perhaps unrelated but interesting
facts about the Ising spin glass partition function zeros. Since the set
of these partition function zeros for finite volumes appears to form a
highly complicated geometrical domain in the complex temperature plane,
it is important to unravel underlying regularities in the distribution.
In particular,
it would be desirable to be able to characterize the set of zeros by
general class properties, as narrowly defined as possible. If we use as
fundamental variable $u \equiv \exp[-2\beta]$ (where $\beta = 1/T$ is the
inverse temperature, generalized to the complex plane), the finite-volume
partition function for an Ising spin glass is, for a fixed random bond
distribution, a finite polynomial in $u$. Instead of working with
the partition function zeros themselves, we may therefore, equivalently,
work directly with this polynomial.

While the
distribution of zeros corresponding to the finite-volume partition function
itself may be difficult to characterize in an exact manner, it may be
possible to find (linear, if possible, but at least invertible)
transformations that take the polynomial of the finite-volume partition
function into new polynomials with zeros located in more simple domains.
Characterizing the distribution of zeros of the transformed polynomials
is equivalent to characterizing the finite-volume partition function zeros
themselves. As examples, we show that certain linear transformations on
polynomials, used in graph theory to classify
graphs \cite{graph}, can be used to transform the partition function
zeros for arbitrary bond distribution into strikingly simpler
geometric domains. For example, we will
give an example of a transformation that, to the numerical accuracy we have
available, maps all zeros of all our available samples onto the real
line. Some of these transformations are described in
section 3. These, or other transformations, may also provide
alternative partial characterizations of the Ising spin glass partition
functions,
although the resulting transformed polynomial may not in itself have
a thermodynamic interpretation as a partition function of a physical
system.

Studying the partition function zeros of an Ising spin glass may on the
surface seem to be a rather isolated problem. In fact, we shall argue on
the contrary. Using the distribution of zeros in the complex plane of
given complicated high-order polynomials (or even distributions of such
polynomials, as in the present case) to classify or ``reconstruct'' various
properties of the polynomials can be a tool of much wider
generality, and may occur in many branches of science. We have already
discussed the fact that precisely the same situation occurs in graph
theory. In spin glass theory it gives a new, at the moment only numerical,
handle with which to
extract physical information about the infinite-volume partition function.
Perhaps it may be amenable to analytical approaches as well.
We hope to have convinced the reader about this point of view by the time
we reach section 4, which contains our conclusions.

\section{Random Polynomials and Spin Glasses}

Consider the $\pm 1$ Ising spin glass,
described by the finite-volume partition function
\beq
{\cal Z}_N(J,\beta) ~=~ \sum_{\{ \sigma \}} \exp\left[- \beta \sum_{i,\mu}
(1 - J_{i,\mu}\sigma_i\sigma_{i+\mu})\right]
\eeq
and the free energy
\beq
F_N(\beta) ~=~ \sum_{\{J \}}\ln[\cal{Z}_N(J,\beta)]/\sum_{\{J\}}
\eeq
Here the sum on $\sigma$ runs over all spin configurations where
$\sigma_i =\pm 1$,
$J$ runs over all bond
configurations with $J_{i,j} = \pm 1$ and $\mu$ denotes the lattice
translations defining the nearest neighbours. In the language of above,
we have $K_{ij} = \beta J_{ij}$. $N$ is the number of sites,
or volume, of the finite lattice defining the system.
It is convenient for the present purposes to rewrite this partition function in
terms of a sum over the energy degeneracies $P_N(E)$, $i.e.$ the number of
states with energy
\beq
E(\sigma ) = \frac{1}{2} \sum_{i,\mu} (1 - J_{i,\mu}\sigma_i\sigma_{i+\mu})
\eeq
on the given finite-size lattice. Then, introducing the variable
$u \equiv \exp[-2\beta]$, we have
\beq
{\cal Z}_N(J,u) ~=~ \sum_{E=0}^{N_b} P_N(E) u^E ~,
\eeq
where $N_b$ is the number of bonds on the lattice.

A crucial observation is that for any finite volume $N$, the relevant
definition
of partition function zeros in the quenched Ising spin glass case must consist
of the
accumulation of partition function zeros for each particular bond
distribution. This follows directly from the definitions (2) and (4),
which show that the quenched free energy
\beq
F_N(u) ~=~ \ln[\prod_{\{J\}} {\cal Z}_N(J,u)]/\sum_{\{J\}}
\eeq
behaves as if it had been obtained from
an effective partition function of product form, once over each
bond configuration.
The effective partition function zeros can thus be computed in a step-by-step
manner, one bond configuration at a time. For practical purposes, a
sampling of configurations should provide enough
information. Lee-Yang zeros and Fisher zeros have previously
been computed numerically using this approach \cite{Ozeki,Bhanot}
for 3-dimensional Ising spin glasses on lattices of varying
finite sizes. Both kinds of partition function zeros hint, with
the statistics available in those references,
at an interesting structure in the infinite-volume limit.

A subtlety and possible weakness in the above line of reasoning ought to be
mentioned at this
point. Formally, in the infinite-volume limit, when {\em all} partition
function zeros of {\em all} bond configurations $\{J\}$ are considered, they
will,
in the above perspective, necessarily lead to physical singularities.
Indeed, consider the ``pure'' Ising bond configuration, one of the
bond configurations included in the infinite sum over $\{J\}$. In the
infinite-volume limit this model has a genuine singularity at a non-zero
temperature for all dimensions larger than 1. Should this singularity,
and its associated partition function zero be of special importance for
the Ising spin glass? Naively, on the basis of eq. (5) the answer would
be yes. But the flaw in this argument lies in the fact that not only does
the free energy contain a piece given by the logarithm of all individual
fixed bond configuration partition functions, it also contains the averaging
over configurations. Clearly the Ising model partition function, and its
associated partition function zeros, will be of ``measure zero'' in the
total sum. But what about nearby models, partition functions with bonds
that are almost pure Ising-like? If the density of partition function
zeros of such models is large enough, it {\em could} lead to a singularity
in the spin glass free energy. Whether it does or not, and where
precisely a genuine singularity occurs in the free energy of the spin
glass thus cannot be settled by looking at individual bond
configurations, and just tracing the location of the associated
partition function zeros. It is
only the total {\em sum} of all zeros that is meaningful, and
here some notion of
regularization is required. Indeed, the above state of
affairs illustrates a potential difficulty with a proper
definition of the spin glass effective partition function
from the limit (as the volume
is sent to infinity) of all bond
configurations in a finite volume, using eq. (5). A perhaps not
unrelated problem is, in more physical language, that as the volume
is increased, there may be new ground states at every new length scale.
In mathematical terms, the strict definition of the infinite-volume spin
glass effective partition function (5)
may require a proper regularization (zeta-function
regularization, or otherwise), and this regularization may shift the
eigenvalues of the transfer matrix away from their naive values.
Basically, what can happen is that the naive location of a zero in the
factorized infinite product (5) may occur at a point in the complex
plane where the remaining infinite product fails to converge. In this
manner a divergence and a potential zero can compensate each other,
producing a displaced zero. This is
a problem inherent to defining the effective spin glass partition function
itself in this manner, instead of relying on the indirect quenched-average
prescription. We will have nothing new to say about it. We will simply
take it for granted that a finite-volume study of this model is meaningful
in the sense that it may yield, by extrapolation, information about the
eventual object of study, the infinite-volume partition function itself.

For a given bond distribution $\{J_{ij}\}$, the finite-volume partition
function ${\cal Z}_N(J,u)$ is a polynomial of degree $n \leq N_b$ in
the variable $u$. The coefficients $P_N(E)$ of this polynomial are
drawn from a probability distribution determined by the random bond
distribution $\{J_{ij}\}$. This is an example of what in the mathematical
literature is known as a {\em random polynomial} \cite{random}. Although
distribution of zeros of various classes of random polynomials with
{\em continuous} coefficient distributions has
been the subject of much study (see, e.g., ref. \cite{random}),
few results are known for random polynomials with discrete
distributions of coefficients, and none for the particular
distribution $P_N(E)$ obtained from
the expansion (4). (For a comprehensive discussion of mathematical
results concerning the location of zeros of polynomials, see also ref.
\cite{geometry}.) Clearly,
we can view our problem from a more general perspective, and it is
tempting at this stage to consider a ``number-theoretic partition
function'' ${\cal Z}_N$
and a ``free energy'' $F_N(\beta)$ by the definitions (4) and (2) for
{\em any} distribution of coefficients $P_N(E)$, not necessarily related to any
statistical mechanics problem. In this context, the index $N$ would rather
denote a given (arbitrary) way of restricting the possible coefficients and
degrees. In particular, one can imagine deforming the Ising spin glass
partitions by choosing
the coefficients $P_N(E)$ from a distribution close, but not exactly equal,
to that of the Ising spin glass. Such generalizations may be of interest
in their own right, but we will consider them here only to get a preliminary
idea of what kind of distribution of partition function zeros we may
expect for the true Ising spin glass. The enormous advantage of
choosing a deformed distribution for $P_N(E)$ is that we can
supply it by hand, thus avoiding the precise determination of the Ising
spin glass
partition function for all given bond distributions. Such
an approach has in fact been suggested earlier by Derrida
\cite{Derrida2} in the special case of a gaussian
distribution, where it is known as the random energy model.
In what follows, we will examine the zeros of related
classes of polynomials whose features make them more and more similar to
an Ising spin glass partition function for a generic bond configuration.
The similarities as well as differences observed at the level of the zeros
will help in understanding the specificity of the Ising glass case.

\section{Distribution of Zeros}

To get some preliminary idea of what kind of distribution we can
expect for random polynomials with positive integer coefficients,
let us first consider the class of
polynomials with a bound on the degree, $N_{max}$, and a common bound
on the coefficients $C_{max}$. Shown in fig. 1 is: (a) a representative
plot of zeros of random polynomials with $N_{max}\!=\!4$ and
$C_{max}\!=\!6$; (b) $N_{max}\!=\!10, C_{max}\!=\!5$; (c)
$N_{max}\!=\!10, C_{max}\!=\!100$; (d) $N_{max}\!=\!30, C_{max}\!=\!5$
(in this last case setting all odd coefficients to zero, for reasons
explained below).

The first remarkable fact is that the zeros are not uniformly spread in
the complex plane, but are centered around the unit circle.
Also, there are clearly defined  zones of much higher density, their number
corresponding exactly to the value of the bound on the degree. Both
features have proven
to be generic, at least in all the cases we have examined. It will
turn out that it takes very particular distributions of coefficients
to deviate from this general pattern. Another remarkable feature is
the presence of holes centered at roots of
unity. Let us study the evolution of this feature as we relax the
bound on the coefficients of the random polynomial.
Figures 1b and 1c show that when we
increase the range of the allowed coefficients, the average
size of the holes tends to decrease, but one can keep on tracing them
on a smaller scale.

In fact, in the very special case of $C_{max}\!=\!1$ \cite{randomzero},
it could be proved that all zeros of all random polynomials within
that particular class are enclosed within a narrow region (the exact definition
of which can be found in ref. \cite{randomzero}) around the unit circle
in the complex plane. The results of that paper are of particular
interest to us, and we shall frequently refer to it in what follows.

Figure 2a shows the situation for $N_{max}\!=\!18$, and figure 2b is a
blown-up region of it, in both cases having $C_{max}\!=\!1$. The
original figures of ref. \cite{randomzero} are of higher quality, but
we include our versions of them in order to be able to compare later
with our corresponding figures for spin glasses. Note that
there is a tendency of the zeros to accumulate on
minute segments of arc-like curves, giving the pictures a highly
non-trivial appearance,
especially at the borders of the set. It gets quickly impractical to obtain a
higher density of points in these regions, as the corresponding polynomials
seem to be fairly rare among all the others. In the work of
Odlyzko and Poonen \cite{randomzero},
the study of these features could nonetheless be pursued much further by
using a reverse procedure which is directly tied with that particular
distribution of coefficients. Instead of explicitly
computing the zeros of polynomials,
the authors systematically tested if, conversely,
a given point in the complex plane could
be a zero of some random polynomial belonging to the case
$C_{max}\!=\!1$, hereafter denoted the 0-1 class. This
was based on a cascade of inequalities peculiar to
the 0-1 case. Working at the maximal resolution of their printer, it enabled
them to provide spectacular evidence that the limiting set of the zeros,
when the degree is arbitrarily large, corresponds to a fractal.
Odlyzko and Poonen could also
provide a heuristic explanation of the self-similarity at
the analytical level, but again, this seems to be possible only because
of the specificity of the 0-1 situation. Nonetheless, we regard these
results as important also for our case, because they illustrate, among others,
how a fractal structure of zeros of a given class of polynomials can emerge
in a {\em practical numerical} study. It was thus important to include
our versions of the Odlyzko-Poonen figures obtained by the brute-force
approach, although we here to great advantage could have used the
reverse procedure. The direct method is the only way we have available
to tackle more complicated classes of polynomials (in particular,
those related to spin glasses).

Of course, we are not expecting
that {\em any}
discrete distribution of coefficients will lead to a fractal distribution of
the zeros. One interesting and relevant counterexample is precisely the
case of the random energy model \cite{Derrida2}. Here the coefficients
$P_N(E)$ in
eq. (4) are chosen according to a Gaussian distribution. This leads to
an essentially solvable model, and interestingly also the distribution of
partition function zeros can be computed exactly \cite{Derrida3}. It turns
out to be a {\em regular} (non-fractal) set in the complex plane,
which also, to the
statistics available, has been observed in numerical analyses
\cite{Moukarzel}. So spin-glass behaviour may certainly not in itself
necessarily imply a fractal set of partition function zeros. With this
perspective in mind, let us now turn to the results of the present
investigation concerning this issue.

\subsection{The Ising spin glass zeros}
We will here present some examples of
the sets of zeros corresponding to genuine Ising glass
partition functions. We have obtained numerous plots of partition
function zeros, for a wide of range of lattice sizes, mostly in 2
dimensions, but also some in 3 dimensions. The figures we have
selected here are the most illustrative of the generic features we
have observed.
Before commenting on the pictures, let us explain how
the partition functions themselves have been obtained.
The method we used has previously been described in
ref. \cite{Bhanot}, so we shall only briefly outline the main
idea (for a different approach, see ref. \cite{Saul}).
For a given bond configuration $\{J_{ij}\}$, the partition
function is evaluated exactly using a numerical transfer
matrix technique \cite{Binder}, which recursively updates the partition
function while building up the $D$-dimensional lattice, stacking
$(D-1)$-dimensional slices one by one. In \cite{Bhanot}, the finite
geometries of the lattices so obtained were chosen to correspond to
``helical boundary conditions'', which minimized the
finite-size errors in the derivation of the corresponding low-temperature
series. In our present case, where global topological properties of the set
of zeros, and its dynamics with increasing $N$, are of primary interest,
we will assume that the boundary conditions are
of second importance. One has only to pay attention to the correct scaling
of the size of the lattice, so that it defines a constantly $D$-dimensional
pattern, and does not degenerate into one of smaller dimension. Only under
these conditions, a finite-size scaling analysis
of the location of the zeros can be meaningfully undertaken. But such
a finite-size scaling analysis will not be performed in our
present study, where
we in fact have mostly used helical lattices.
They can be defined in the following way
(for a more complete treatment, the reader is urged to consult ref.
\cite{helical}) . Consider sites numbered sequentially
with integers, and, for a system of dimension
$D$, a set of $D$ integer ``periods'': $h_1\leq h_2\leq ...\leq h_D$, so that
the nearest neighbours of a site $i$ in the direction $j$ are defined by the
jumps $i \pm h_j$. The biggest period $h_D$ can be viewed as defining one
``turn'' of a helix. The total helical (finite) lattice is then built up
from superposing the desired number of such turns. The corresponding
partition function (for a given bond distribution $J$) is obtained by
the recursive procedure described above. Helical lattices of this type are
locally hypercubic, but their global geometry is topologically more involved.
Their resulting behaviour, as far as distribution of zeros is
concerned, should,
however, be qualitatively similar to those of more conventional periodic
hypercubic lattices.

We have mainly considered 2-dimensional models where we could more
efficiently gather a substantial number of data points. However,
consideration of some 3-d cases showed that they
exhibit similar features at the
level of their distribution of zeros. Typically, the number of points
per plot is of the order of 100,000 (which was also the case for the
previous figures). Our algorithm checked that the zeros generated by
one random bond configuration did not reproduce the zeros already obtained.
In our lattices, each site is connected to 2d neighbours, which causes
all $P_N(E)$ with $E$ odd to vanish, as can be easily proved. The resulting
distributions of zeros have consequently symmetry both about
the real and the imaginary axis.
It is thus sufficient to examine only what happens in the first
quadrant of the complex plane. Figures 3a-d all deal with the 2-d
Ising spin glasses. Lattice sizes are (a) $N\!=\!18, h_2\!=\!7$, (b)
$N\!=\!18, h_2\!=\!3$, (c) $N\!=\!18, h_2\!=\!3$, (d) $N\!=\!18,
h_2\!=\!6$, and (e) $N\!=\!19, h_2\!=\!3$,
the figure 3c showing finer details of 3b.
The very first striking fact is the dissimilarity with the figures 1
and 2.
Instead of a high concentration on the unit circle, and almost void interior
region, the contrary seems to occur here. This is troubling, because,
taking the bound on the coefficients sufficiently high, the flat distributions
of the preceding section are generating, among others,
all genuine Ising glass
polynomials. It seems then that the latter must be particularly scarce.
Indeed, for a given volume $N$, the maximal degree of the spin glass
partition function is $N_b\!=\!2N$, and the bound on the coefficients
is certainly grossly overshot by $2^N$. There are then $2^{N(N_b+1)}$
random polynomials, which contain in particular the $2^{2N}$ spin
glass partition functions. However, this does not explain why the
zeros of the Ising spin glass partition functions do not tend at all
to accumulate around the unit circle, despite the fact that this
generically is the most dense region.
To understand the situation better, one can first examine the effect of
setting to zero, in the flat case, all the odd coefficients. Figure 1d
already
showed that this restriction does not explain the dissimilarity.
If one examines the Ising glass partition functions, another feature is
apparent. For any bond distribution, there seems to exist an index $E_0$
such that $P_N(E) < P_N(\bar{E})$ for $E < \bar{E} < E_0$ and
$P_N(E) > P_N(\bar{E})$ for $E > \bar{E} > E_0$, where
only non-vanishing coefficients are considered. Sequences of integers with
this property are well known in combinatorics and graph theory where it is
called unimodality. In all configurations we have studied, we have
never encountered any violation of unimodality, but we have no direct
proof that in fact all Ising spin glass distributions have this
property.

The next obvious step is then to
deform the flat distributions considered above in such a manner that
unimodality is incorporated. Figure 4a  shows the result
for polynomials with $N_{max}\!=\!30, C_{max}\!=\!100,000$ where all
odd coefficients are set to zero, and where unimodality instead is
being implemented in the following manner. We take, for a random
number $r$ flatly distributed between 0 and 1, the remaining even
coefficients  $N_{k+2} = N_kr(C_{max}- N_k)$ up to the ``middle''
coefficient. Past this point we take $N_{k+2} = N_kr$.
We see that we get closer to the pictures of figure 3, albeit this does not
seem to
be the end of the story. To more drastically remove zeros from the
unit circle, and place them closer to the interior region, consider
figure 4b, which show the corresponding zeros with $N_{k+2}=
(N_kr)^{\alpha}$ ($\alpha$ being a flatly distributed random number
between 1 and 2) up to the middle of the polynomial, the procedure
being mirrored for the remaining $N_k$'s, i.e. such that
\beq
N_{N_{max}-2k} ~=~ N_{2k} ~.
\eeq
It looks as if this exponential growth of the coefficients better
``simulates'' the Ising spin glass case. But clearly other
``correlation'' effects among the coefficients,
much harder to characterize, are important as well.

It can be
shown that for helical lattices with $N$ even, $h_1\!=\!1$ and $h_2$
odd, for each spin configuration of energy $E$, there exists another
configuration of energy $N_b-E$. This means that the corresponding
partition function has the mirror symmetry
$P_N(E) = P_N(N_b-E)$. As a
consequence, if $x_0$ is a zero, then so is $1/x_0$. This feature is
not generic, and of course does not explain the suppression of zeros
around the unit circle. Furthermore, the structure of the fine details
of the zeros (and particularly around the boundaries) is {\em not}
affected by imposing this mirror symmetry, cf. figure 3d, which is
not mirror-symmetric.

Apart from these global aspects, a closer look at the zeros
of actual Ising spin glass partition functions reveals that, remarkably,
many of the small-scale
features of the 0-1 case \cite{randomzero} are present here too.
The zeros tend to coagulate, and, at the boundaries,
the same ``spike'' or ``cusp''-like organization seems to take place. This
is especially evident in figures 3c and 3d, which display distinct
fractal-like features along the boundary.
Although it is far from conclusive, the similarity
of these boundary features with the ones present in the known
fractal case of \cite{randomzero} (compare figure 2) certainly
makes the hypothesis that the Ising spin glass partition function
zeros may accumulate on a fractal set a genuine possibility.

\section{Graph Theory and Polynomial Mappings}

As it turns out, some of the mathematical machinery used in graph
theory to classify graphs can with advantage be applied to Ising spin
glass theory as well. Although there seems to be no direct mapping
between finite-volume Ising spin glass partition functions and particular
graphs, certain transformations of polynomials used in graph theory
may be useful here as well.

Since most of the notions we will be borrowing from graph theory are not
readily available in the physics literature, we begin with a
few definitions.\footnote{Most of the presentation below follows the
excellent exposition in ref. \cite{Brenti}.}
In graph theory, one can associate to each graph $G$ a polynomial,
the {\em chromatic polynomial} $P(G;x)$, first introduced by Birkhoff.
We do not need its precise definition in terms of a given graph (for a good
introduction to the subject, see ref.\cite{graph}), but it is important
that it can be represented by an expansion
\beq
P(G;x) = \sum_{j=0}^n a_j(G)(x)_j ~,
\eeq
with non-negative {\em integer} coefficients $a_i$. Here,
\beq
(x)_j \equiv x(x-1)\cdots(x-j+1)
\eeq
is known as the $j$-th {\em falling factorial} polynomial, and $n$ is
the number of vertices of the graph. The coefficients $a_j$ of the
expansion (7) have a direct interpretation in terms of partitions of
vertices of the graph; $a_j$ is the number of inequivalent way of
dividing the vertices of
the graph $G$ into exactly $j$ blocks, each inducing what in graph
theory is known as an edge-free subgraph of $G$ (see ref. \cite{graph}).
It is then easy to see that, for an integer $k$, $P(G,k)$ gives the number
of proper colourings of the graph $G$ using $k$ colours.

One important property of the chromatic polynomial is related to its
expansion in the $x^j$-basis instead of the $(x)_j$-basis used in the
definition (7) above. It is readily shown that in this basis the
coefficients are alternating in sign, so that we may introduce non-negative
integer coefficients $b_j$ defined by
\beq
P(G;x) = \sum_{j=0}^n (-1)^{n-j} b_j(G)x^j ~.
\eeq
The coefficients $b_j$ have been conjectured to be not only {\em unimodal},
\beq
b_0 \leq \cdots \leq b_k \geq \cdots b_n ~,
\eeq
for some index $k$ (with $0 \leq k \leq n$), but in fact {\em strictly
logarithmically concave}:
\beq
b_j^2 ~>~ b_{j-1}b_{j+1} ~.
\eeq

It is precisely this conjecture which, among others, motivates the study of the
zeros of the chromatic polynomials.
Also, the location of the zeros of the polynomial can be used as a
criterion in the ``reconstruction'' or ``inverse'' problem which consists
in examining the conditions upon which a given polynomial can be
a chromatic polynomial of some graph. Although necessary conditions are
relatively
easy to derive (see \cite{graph}), sufficient ones are not known in general.
Researchers in this field are therefore confronted with problems which are
quite similar
in nature to those met in our present study. In general a class of
polynomials obtained by an explicit process is studied from the point of
view of their zeros, but the process itself gets quickly
too complex to enable a direct investigation of the general case.The only
possibility is then to remain within
small volumes or, in the language of graph theory, a small number of
vertices. Here computer algorithms can be used.
One may therefore hope that techniques which have
proven useful in one field may be
successful in the other as well. What follows is devoted to examining one
such possibility, based on recent results in \cite{Brenti}.

The zeros of general chromatic polynomials, although bounded in several
ways in the complex plane, fall in highly irregular regions. But
certain associated polynomials, directly derivable from the chromatic
polynomial itself, have zeros that display striking regularities.
Given a chromatic polynomial $P(G;x)$, one can introduce three related
polynomials that are particularly useful: the $\sigma$-polynomial, the
$\tau$-polynomial, and the
$\omega$-polynomial. These are defined as follows \cite{Brenti}. The
$\sigma$-polynomial
is obtained from the chromatic polynomial by taking the number of partitions
$a_j(G)$ of eq. (7) as fixed coefficients, but replacing the basis
$(x)_j$ by the basis $x^j$. Thus,
\beq
\sigma(G;x) = \sum_{j=0}^n a_j(G) x^j ~,
\eeq
which obviously should not be confused with the expansion of the chromatic
polynomial itself in the $x^j$-basis, eq. (9).

To construct the $\tau$-polynomial, first expand the
chromatic polynomial $P(G;x)$ in the $\langle x\rangle_j$-basis as well:
\beq
P(G;x) = \sum_{j=0}^n (-1)^{n-j}c_j(G)\langle x\rangle_j ~,
\eeq
where
\beq
\langle x\rangle_j \equiv x(x+1)\cdots(x+j-1)
\eeq
is the $j$-th {\em rising factorial} polynomial. Then define the
$\tau$-polynomial by taking the coefficients $(-1)^{n-j}c_j$, but replacing
the basis $\langle x\rangle_j$ by $x^j$, $viz.$,
\beq
\tau(G;x) = \sum_{j=0}^n (-1)^{n-j}c_j(G) x^j ~.
\eeq

Finally, introduce the $\omega$-polynomial by
\beq
\omega(G;x) = \sum_{j=0}^n h_j(G) x^j = (1-x)^{n+1}\sum_{m} P(G;m)x^m ~,
\eeq
where the sum in the last expression runs over all natural numbers. The last
identity in eq. (16) is highly non-trivial, but not very useful from a
practical point of view. Fortunately, a number of identities exist, which
relate the different polynomials to each other and to more manageable
expressions. For example, one can show \cite{Brenti} that
\beq
\omega(G;x) = (1-x)^n \sum_{j=0}^n j! a_j(G)\left(\frac{x}{1-x}\right)^j ~.
\eeq
This last identity hints at the usefulness of introducing what are
called {\em augmented} $\sigma$ and $\tau$ polynomials, which will be
denoted by $\bar{\sigma}(G;x)$ and $\bar{\tau}(G;x)$, respectively:
\beq
\bar{\sigma}(G;x) = \sum_{j=0}^n j! a_j(G) x^j ~,
\eeq
and
\beq
\bar{\tau}(G;x) = \sum_{j=0}^n (-1)^{n-j} j! c_j(G) x^j ~.
\eeq
In particular, note that
\beq
\omega(G;x) = x(1-x)^n\bar{\tau}\left(G;\frac{1}{1-x}\right)~.
\eeq
There are a number of interesting functional identities between the different
polynomials, and the original chromatic polynomial. They can be useful
because they may establish connections
between the original finite-volume
Ising spin glass partition function and some of the
new polynomials derived from it.

The four polynomials $P(G;x), \sigma(G;x), \tau(G;x)$ and $\omega(G;x)$
can be used to generate an interesting hierarchy of conditions regarding
``reality'' of the polynomials (a polynomial is defined to be real if
all its roots are real) \cite{Brenti}. These reality conditions are of an
entirely general nature, and can hence be used for arbitrary polynomials,
including the ones that have a physical meaning as Ising spin glass partition
functions in a finite volume. To quote some examples: $\omega$-reality
implies both $\sigma$-reality and $\tau$-reality. P-reality implies
$\tau$-reality, and the conditions P-reality, $\tau$-reality,
$\omega$-reality and $\sigma$-reality all imply that the corresponding
P, $\tau$, $\omega$, and $\sigma$ polynomials have coefficients that form
a strictly logarithmically concave sequence. Of course, for polynomials
based on both graph
theory chromatic polynomials and finite-volume Ising spin glass
partition functions there may be further specific relations that are valid
only within these subclasses.

To demonstrate how these mappings can be used in our Ising spin glass
case, consider identifying
\beq
P_N(E) ~=~ a_E ~,
\eeq
so that the actual partition becomes the $\sigma$-polynomial (of
course with only even coefficients, so there is certainly no graph
corresponding to it). Let us now compute the zeros of the associated
$P$-polynomial. This is shown in figure 5a, for a lattice of $N\!=\!12,
h_1\!=\!1, h_2\!=\!7$. As another example, consider the
$\bar{\tau}$-polynomial derived again from the same identification; we
show the zeros of this polynomial for $N\!=\!17, h_1\!=\!1, h_2\!=\!5$
in figure 5b. (This is particularly interesting, because the pure
Ising model treated in the same manner yields a set of zeros that
appears to form a perfect ellipse.) Although we are not showing it
(because it has no structure whatsoever), perhaps the most interesting
plot is that of the zeros of the
the $\tau$-polynomial itself. Here, for all Ising spin
glass partition functions we have considered, {\em all} zeros fall {\em
exactly} on the real axis. The Ising spin glass glass partition
functions thus appear, to the extent we have been able to sample them,
to be $\bar{\tau}$-real.\footnote{An interesting check concerns the
application of these polynomial maps to an exactly solvable case such
as the ordinary 1-d Ising model. Using explicit representations of the
finite-volume model (see, e.g., ref. \cite{Heller}), we have confirmed
that the properties of the mapped polynomials are shared by this
particular model. These mappings may be of use also for the study of
higher-dimensional Ising models.}

We consider the transformations of polynomials described above as just
examples of what might be useful tools for analyzing the original partition
function zeros. The polynomials we have considered here do have a number
of almost magical properties when defined from an original polynomial
with integer (or rational) coefficients, and in that sense several of
the properties discussed above are of a far more general nature, and are not
restricted to random polynomials of distributions corresponding to Ising
spin glasses. Ideally, one should find transformations such that the
transformed polynomial satisfies a certain criterion concerning its zeros
(such as reality) {\em only} in the class of random polynomials that
correspond to Ising spin glass partition functions. Evidently, the broader
the class,
the less suited it will be for a classification of the Ising spin glass
partition functions. But the examples we have given above certainly do
share a number of useful properties, and it is not unlikely that new
related transformations can be used to further limit the class of polynomials
for which the zeros form simple patterns in the complex plane.

\section{Conclusion}

We hope to have convinced the reader at this point that an analysis of
the distribution of zeros of Ising spin glass partition functions is
important. This is a subject of study which has become possible
within the last few years due to the increased
computational power available, and it
has revealed a number of very interesting facts about the Ising spin
glass partition function. At a detailed numerical level, the approach
towards the real temperature axis can give information about the
nature of the spin glass phase transition \cite{Bhanot}, and about its
critical exponents. In this paper we have argued that {\em global}
aspects of the distribution of zeros can contain highly non-trivial
information as well. Although all of our analysis is numerical at this
point, we have presented visual evidence that the full
distribution of partition function zeros is forming a highly
non-trivial set in the complex temperature plane. Whether this
set actually is fractal cannot be proved at this level, but it clearly
remains a genuine possibility. If correct, this would indirectly
provide an intriguing hint that the underlying renormalization-group
dynamics may be chaotic, a connection so far only established in the
perhaps more contrived spin glass models defined by spin systems with
frustration on hierarchical lattices.

We have tried to show that there may be more analytical means of
studying this problem as well. In particular, finding appropriate
polynomial transformations may be the tool with which the highly
complicated distribution of partition function zeros can be brought
under much tighter control. It is plausible that a very specific
polynomial mapping exists, which uniquely selects out Ising spin glass
partition functions as those members of a class of random polynomials
whose zeros, when considered in the transformed basis, fall on very
simple domains (such as the real line). In section 4 we gave various
examples of candidates for such polynomial mappings, thereby making a
(perhaps fortuitous) connection to Graph Theory. These mappings fail,
however, to be sufficiently selective for our purposes.

Instead of relying on visual evidence, a direct way to test whether
the set of partition function zeros of the Ising spin glass fall on a set
of fractal dimension is of course
to compute this dimension on the basis of our
data. Such a procedure is, however, not free of ambiguities. First, the
notion of a fractal dimension is in itself not unique (there are infinitely
many ways of ``analytically continuing'' the common definition of integer
dimensions to the set of reals), and second, extracting the fractal
dimensions in this way on the basis of raw data alone is highly non-trivial.
The non-uniqueness of the notion of a fractal dimension is often, arbitrarily,
parametrized within one given class of fractal dimensions in terms of one
real number $q$. This definition, which dates back to the work on the entropy
of probability distributions by R\'{e}nyi, starts by subdividing the space
(in this case the complex temperature plane) into a more and more fine grid
of linear size $r$. Let $p_k$ denote the probability that an element of the
set is contained within the $k$'th cell. The generalized fractal dimensions
$D_q$ is then defined by
\beq
D_q ~\equiv~ \lim_{r \to 0}\frac{1}{q-1}\frac{\ln\left[\sum_k
p^q_k\right]}{\ln(r)}
\eeq
for all $q \neq 1$. For $q\!=\!1$, the analogous definition is
\beq
D_1 ~\equiv~ - \lim_{r \to 0}\frac{p_k \ln(p_k)}{\ln(1/r)} ~.
\eeq
This latter, $D_1$, is known as the information dimension. $D_0$ is the
more commonly known Hausdorff dimension, and $D_2$ is denoted the correlation
dimension in the literature.

While it is not obvious which fractal dimensions $D_q$ (or others) one should
focus on, a more practical problem concerns the actual determination of
these numbers $D_q$ from our raw data. Although several of our samples of
partition function zeros contain as many as 100,000 points, this number
is in fact not sufficient to determine reliably the limit $r\!\to\!0$ in
eq. (22). The subdivision of intervals leads to an exponential growth in
cells, which quickly clashes with the requirement that the typical number
of points per filled cell should be
larger than one. Thus one quickly reaches a
regime (as a function of subdivisions of the cells) where one is only
measuring the thinning-out of points due to the finite sample. We have
attempted to calculate both the standard Hausdorff dimension $D_0$ and
the correlation dimension $D_2$ in this brute-force manner, but will not
quote any numbers, since they are too inconclusive. The only statement
we can make is that a Hausdorff dimension of two for the {\em interior}
of the set is not incompatible with our results.

How can the issues brought up by the present paper be studied in
greater detail? There are severe numerical limitations in going
significantly beyond the number of computed partition function zeros
(or, equivalently, the lattice sizes that we have reached). Increasing
the number of data points 10-fold, perhaps 100-fold, may be what is
required to really see a possible conclusive fractal structure
emerging, as the 0-1 case discussed in section 3 has indicated.
It may even be that, deviously, an apparent fractal behaviour seems to emerge
on smaller systems, but disappears in the limit of the volume going to
infinity.
Clearly, the converse procedure, an algorithm that can tell (to any
given accuracy) whether a point in the complex plane can belong to the
zeros of Ising spin glass partition functions or not would therefore
be highly advantageous.
Finding such an algorithm is, however, {\em far} more
difficult than the 0-1 case referred to above. In fact, the
availability of such an algorithm would almost amount to having an
explicit solution for Ising spin glass partition functions of
arbitrarily large volumes.

Another aspect that deserves further study in the light of our
findings, is the distribution of Lee-Yang zeros in the complex
activity plane. The very preliminary results reported in ref.
\cite{Ozeki} certainly hint at an analogous structure present in the
set of those zeros. With present-day computers this aspect of spin
glass theory can be pushed much beyond the results known so far.

\vspace{0.5cm}

\noindent
{\sc Acknowledgement:} ~We would like to thank G. Bhanot for a number of
helpful conversations at the early stages of this work, and G. D'Alessandro
for providing us with an alternative code for determining fractal
dimensions. One of us (J.L.)
would like to thank A.M. Odlyzko for very stimulating conversations
about the distribution of zeros of random polynomiasls. He also
thanks B. Derrida for discussions.

\vspace{0.5cm}

\noindent
{\bf Note Added:} ~After the completion of this paper, we became
aware of the recent work by Baake, Grimm and Pisani \cite{Baake}. These
authors demonstrate that a spin model defined on a regular lattice can lead to
a fractal distribution of partition function zeros. To our knowledge,
this is teh first time it has been demonstrated that fractal sets of partition
function zeros can arise in models other than those defined on hierarchical
lattices. The model they consider is a one-dimensional Ising-like theory with
fixed nearest-neighbour couplings that are distributed according to a
Fibonacci sequence.


\vspace{1.5cm}

\begin{center}

\Large{Figure Captions}

\end{center}

\noindent
{\bf Fig. 1a)} Distribution of zeros of random polynomials of degree 4, with
all coefficients taken randomly from a flat distribution between 1 and
6. There is a slight increase in the density of points around four
spots close to the unit circle. There are also distinct holes in the
distribution close to points on the unit circle.

\noindent
{\bf Fig. 1b)} Same as fig. 1a, but this time the degree of the random
polynomial is increased to 10, and the coefficients are sampled
randomly between 1 and 5. There are ten clear regions along the unit
circle where the density is markedly larger. The holes have shrunk in
size.

\noindent
{\bf Fig. 1c)}~ The degree is still 10, but the range of the
coefficients is now increased to the interval [1,100]. The ten regions
of higher density along the unit circle clearly remain, but the holes
have decreased in size.

\noindent
{\bf Fig. 1d)}~ The random polynomial is taken with only {\em even}
powers. We show here an example of degree 30, with coefficients taken
randomly in the interval [1,5]. The
number of regions of higher density is
fourteen, one less than the generic fifteen for a random
polynomial of degree 30. The holes along the unit circle can still be
seen.

\noindent
{\bf Fig. 2a)}~ Zeros in the upper complex plane of random polynomials
of degree 18 and coefficients 0 or 1.

\noindent
{\bf Fig. 2b)}~ Finer details of the fig. 2a-case. The distribution is
believed to be fractal \cite{randomzero}.

\noindent
{\bf Fig. 3a)}~ Partition function zeros of a 2-d Ising spin glass of
the form discussed in the text. Here $N\!=\!18, h_1\!=\!1, h_2\!=\!7$.

\noindent
{\bf Fig. 3b)}~ Same as in fig. 3a, but with $h_2\!=\!3$.

\noindent
{\bf Fig. 3c)}~ Blown-up picture of the partition
function zeros displayed in fig. 3b.

\noindent
{\bf Fig. 3d)}~ Partition function zeros when for the same lattice
above we set $h_2\!=\!6$. For this lattice size there is no reflection
symmetry with respect to the unit circle (such that when $x_0$ is a
root, so is $1/x_0$). The fine details of the distribution of zeros is, as
would be expected, unaffected by this symmetry.

\noindent
{\bf Fig. 3e)}~ Details of the distribution of zeros for a slightly
larger lattice with $N\!=\!19, h_2\!=\!3$, which again does not have
symmetry with respect to inversions around the unit circle.

\noindent
{\bf Fig. 4a)}~ In order to mimic the basic features of the spin glass
partition function zeros, we here consider random polynomials with a
distribution of coefficients of linear growth (until the middle of the
polynomial, and a linear decrease beyond the middle), as explained in
detail in the main text. It is clear that the generic tendency of the
zeros of random polynomials to accumulate close to the unit circle
has been modified. There is a shift toward the interior of the unit
circle. The polynomials are of degree 30, with a distribution of
coefficients between 1 and 100.000.

\noindent
{\bf Fig. 4b)}~ Same as fig. 4a, but this time choosing an even
steeper rise in coefficients until the middle of the polynomial, and a
corresponding steeper decrease beyond the middle. The exact
distribution is described in the main text. Note that the zeros have
been shifted completely away from the unit circle, mimicking many of
the gross features of the genuine Ising spin glass case.

\noindent
{\bf Fig. 5a)}~ Example of how the zeros of the Ising spin glass
partition function look in a ``transformed basis'', as described in
the main text. Shown here is the distribution of the zeros of the
associated $P$-polynomial (the analogue of the chromatic polynomial in
graph theory). The original Ising spin glass lattice corresponds to
$N\!=\!12$.

\noindent
{\bf Fig. 5b)}~ Ising spin glass partition function zeros in the
$\bar{\tau}$-picture, here for a lattice of $N\!=\!17$.

\end{document}